\pdfoutput=1

\documentclass[twocolumn,english]{revtex4}
\usepackage{times,amssymb,amsmath,graphicx}
\usepackage[bookmarks=false,pdffitwindow=false,pdfstartview={FitH}]{hyperref}
\usepackage[T1]{fontenc}
\usepackage{babel,multirow}
\usepackage{color}

\begin{document}


\title{
  Mesoscopic Josephson effect in graphene disk at magnetic field 
}

\author{Adam Rycerz\footnote{Correspondence: 
  \href{mailto:rycerz@th.if.uj.edu.pl}{rycerz@th.if.uj.edu.pl}.}}
\affiliation{Institute for Theoretical Physics,
  Jagiellonian University, \L{}ojasiewicza 11, PL--30348 Krak\'{o}w, Poland}

\date{April 25, 2026}

\begin{abstract}
Unlike for tunneling Josephson junctions, for which the current-phase
relation is given by the sine function, with the critical current ($I_c$) and
normal-state resistance ($R_N$) following the relation
$I_cR_N=(\pi/2)\,\Delta_0/e$ (where $\Delta_0$ is the superconducting gap and
electron charge is $-e$), mesoscopic Josephson junctions
show more complex current-phase relations, with the skewness $S>0$,
what is related to the presence
--- in case the leads are in the normal state --- of transmission
probabilities  taking the values comparable to $1$.
Here, we show that these features also appear for 
a~superconductor-graphene-superconductor (S-g-S) junction in the disk-shaped
(Corbino) geometry, when the magnetic field is adjusted such that
$I_c\rightarrow{}0$ and $R_N\rightarrow{}\infty$.
In such a~case, the product $I_cR_N\approx{}1.85\,\Delta_0/e$, and the skewness
$S\approx{}0.14$. 
The results obtained from quantum-mechanical mode-matching analysis
for the Dirac-Bogoliubov-De-Gennes equation are
compared with simpler model assuming incoherent scattering 
between two circular interfaces separating the sample and the leads. 
\end{abstract}

\maketitle


\section{Introduction}
The Josephson effect, in which a~dissipationless (superconducting) current
passes between two superconductors via an insulator \cite{Jos64}
(or, more generally, via a~normal metal \cite{Vol95,Mor98}, a~semiconductor
\cite{Cla80}, or a~nanostructure \cite{Doh05,Sim19}) has gained
a~renewed attention in the context of quantum computations (see, e.g.,
Ref.\ \cite{Wen17}). 
The significance of this area of research
was recognized with the Nobel Prize in Physics 2025 for Clarke,
Devoret, and Martinis, who demonstrated the occurrence of macroscopic quantum
effects in Josephson junctions coupled to a~microwave field in the mid-1980s 
\cite{Mar85,Dev85}. 
After the fabrication of semiconducting heterostructures with the
two-dimensional electron gas (2DEG) \cite{Din78,Mim80}, an interest in
superconductor-2DEG-superconductor Josephson junctions raised
\cite{Tak85,Aka94}, mainly because such systems allow one to control
the essential junction parameter --- the critical
current $I_c$ --- by external electrostatic rather than magnetic field
\cite{Gol21}. 

The advent of graphene \cite{Kat20} motivated researchers to reexamine the
Josephson effect for superconductor-graphene-superconductor (S-g-S) junctions
\cite{Tit06,Tit07,Hee07,Eng16,Nan17}. 
In analogy to early studies of the Josephson effect, theory \cite{Tit06,Tit07}
preceded experiments \cite{Hee07,Eng16,Nan17}, both showing that S-g-S
junctions have intrinsic (graphene-specific) features, including the amplified
product of the critical current and normal-state resistance $I_cR_N$, and
forward skewness of the current-phase relation.

\begin{figure}[b]
\includegraphics[width=\linewidth]{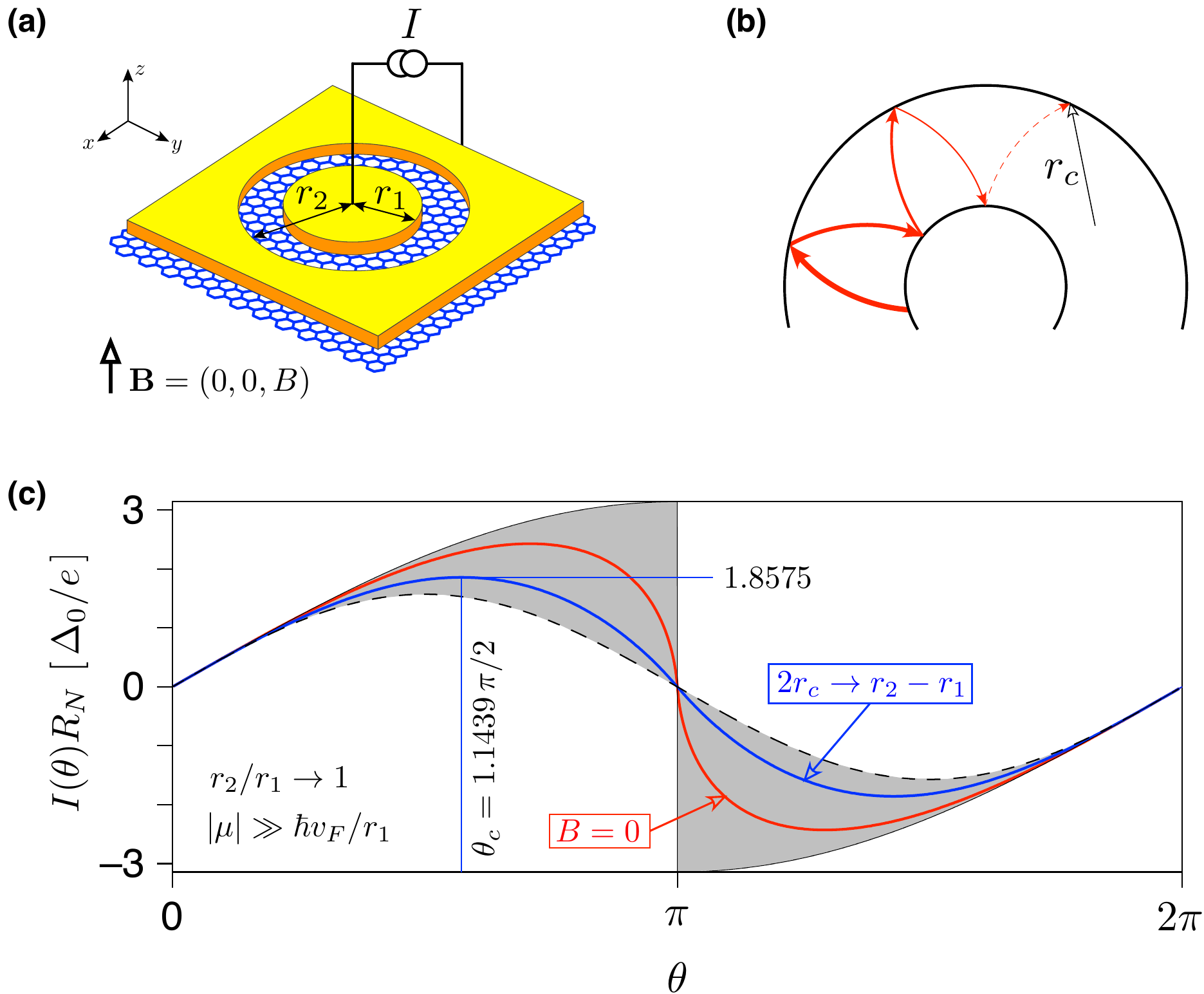}
\caption{ \label{setupfig}
(a) Schematic of a~graphene disk with inner radius $r_1$ and outer radius
$r_2$, contacted by two circular superconducting electrodes (yellow areas).
A current source drives a dissipationless supercurrent through the annular
region. The uniform magnetic field $\mathbf{B}=(0,0,B)$ is
perpendicular to the sample. A separate gate electrode (not shown) allows
one to tune the carrier concentration. 
(b) At low field but high doping, such that the cyclotron diameter
$2r_c>r_2-r_1$, incoherent scattering along the classical trajectory governs
the transport.
(c) Current-phase relation for high doping ($|\mu|\gg{}\hbar{}v_F/r_1$)
and the narrow-disk limit ($r_2/r_1\rightarrow{}1$).
Red line correspond to the $B=0$ case (see Ref.\ \cite{Ryc26b}),
blue line visualizes the $2r_c\rightarrow{}r_2-r_1$ limit studied in this
work. Thin black lines mark the tunneling limit (dashed)
and the ballistic limit (solid). 
}
\end{figure}

In this paper, we consider the Corbino geometry to point out that S-g-S
junction in magnetic field still show graphene-specific features even in the
limit of vanishing conductance (i.e., when the cyclotron diameter approaches
the distance between superconducting electrodes, see Fig.\ \ref{setupfig}),
albeit one can expect the tunneling behavior in such a~case. 

The paper is organized as follows.
Essentials of the Dirac-Bogoliubov-De-Gennes theory, with the exact formula
for transmission probabilities for the Corbino-Josephson setup, are given
in Sec.\ \ref{modmet}.
Next, the approximation assuming incoherent scattering between two interfaces
is presented in Sec.\ \ref{incoscat}.
Our main results, concerning the critical current, normal-state resistance,
and skewness of the current-phase relation for different dopings and magnetic
fields are described in Sec.\ \ref{resdis}. 
The conclusions are given in Sec.\ \ref{conclu}.

\section{Theory and definitions}
\label{modmet}
In the framework of Dirac-Bogoliubov-De-Gennes theory, measurable quantities
for an S-g-S Josephson junction are determined by the transmission
probabilities ($T_j$), which can be calculated for infinitely-doped graphene
leads in the normal state \cite{Tit06,Tit07}. 
For the Corbino geometry and the magnetic field $B>0$ \cite{Ryc10}, 
see Fig.\ \ref{setupfig}, 
\begin{equation}
\label{tjbnon}
  T_j=|t_j|^2=
  \frac{16\,(k^2/\beta)^{|2j-1|}}{k^2{}r_1{}r_2{}%
  \,(X_j^2+Y_j^2)}
  \left[\frac{\Gamma(\gamma_{j\uparrow})}{\Gamma(\alpha_{j\uparrow})}\right]^2,
\end{equation}
where $j=\pm{}\frac{1}{2},\pm{}\frac{3}{2},\dots$ is the angular-momentum
quantum number, $k=|\mu/(\hbar{}v_F)|$, $\beta=eB/(2\hbar)$, 
$\Gamma(z)$ is the Euler Gamma function, $\gamma_{js}=|l_s|+1$ and 
$\alpha_{js}=\frac{1}{4}[2(l_{-s}+|l_s|+1)-k^2/\beta]$ with
$l_s=j\mp\frac{1}{2}$ for $s=\uparrow,\downarrow$. 
\begin{align}
& X_j = w_{j\uparrow\uparrow}^- + z_{j,1}z_{j,2}w_{j\downarrow\downarrow}^-,\ \ \ 
  Y_j = z_{j,2}w_{j\uparrow\downarrow}^+ - z_{j,1}w_{j\downarrow\uparrow}^+, \nonumber\\
& w_{jss'}^\pm= \xi_{js}^{(1)}(R_\mathrm{i})\xi_{js'}^{(2)}(R_\mathrm{o})
  \pm \xi_{js}^{(1)}(R_\mathrm{o})\xi_{js'}^{(2)}(R_\mathrm{i}),  
\end{align}
where $z_{j,1}=[2(j+s_j)]^{-2s_j}$, $z_{j,2}=2(\beta/k^2)^{s_j+1/2}$, with 
$s_j\equiv\frac{1}{2}\mbox{sgn}\,j$ (${\rm sgn}(x)$ is the sign function),
and 
\begin{equation}
\label{xisnu}
  \xi_{js}^{(\nu)}=e^{-\beta{r}^2/2}(kr)^{|l_s|}\left\{\begin{array}{cc}
    M(\alpha_{js},\gamma_{js},\beta{r}^2), & \nu\!=\!1, \\ 
    U(\alpha_{js},\gamma_{js},\beta{r}^2), & \nu\!=\!2. \\
  \end{array}\right.
\end{equation}
$M(a,b,z)$ and $U(a,b,z)$ in Eq.\ (\ref{xisnu}) are the confluent
hypergeometric functions \cite{Abram}. 
For $B<0$, it can be shown that $T_j(B)=T_{-j}(-B)$. 

At zero temperature and for the short-junction limit, i.e., 
$r_2-r_1\ll{}\xi_0$ (with the superconducting coherence length
$\xi_0=\hbar{}v_F/\Delta_0$), the Josephson current can be written as
\begin{equation}
\label{ijoth}
  I(\theta)=\frac{e\Delta_0}{\hbar}\sum_{j}
  \frac{T_j\sin\theta}{\sqrt{1-T_j\sin^2(\theta/2)}},  
\end{equation}
while the normal-state resistance is given by
\begin{equation}
\label{rnlan}
  R_N^{-1}=\frac{4e^2}{h}\sum_{j}T_j.
\end{equation}
The remaining symbols are $\Delta_0$ --- the superconducting gap, and
$v_F$ --- the energy-independent Fermi velocity in graphene.
(In the physical units, $\hbar{}v_F=0.575214\,$eV$\cdot$nm.) 
For instance, $\xi_0\approx{}550\,$nm for superconducting
electrodes made with molybdenum rhenium \cite{Nan17}.

Although the physical quantities we consider depend only on dimensionless 
parameters, namely
\begin{equation}
\label{unitdefs}
  a=\frac{r_1}{r_2},
  \ \ \ \ \
  \mu^\star = \frac{\mu{}r_1}{\hbar{}v_F},
  \ \ \ \ \
  B^\star = \frac{eBr_1^2}{2\hbar},  
\end{equation}
one can expect the short-junction approximation used in Eq.\ (\ref{ijoth})
to be applicable up to about $r_2-r_1\lesssim{}200\,$nm \cite{unitfoo}.
Additionally, the superconducting circular leads introduce the flux
quantization, and the average field in the sample area is an integer
multiplicity of $\Phi_0/\pi{}(r_2^2-r_1^2)$, with $\Phi_0=\pi{}\hbar/e=
2067.83\,$T$\cdot$nm$^2$. 
Using the units introduced in Eq.\ (\ref{unitdefs}), we can write
\begin{equation}
\label{bsquant}
  B^\star = \frac{a^2}{2(1-a^2)}\,l,\ \ \ \ 
  \text{with }\ l=0,1,2,\dots. 
\end{equation} 

It is worth pointing out that the current-phase relation $I(\theta)$ can
be expressed via the charge-transfer cumulants in the normal state.
Namely, expanding $(\sqrt{1-x})^{-1}$ with respect to $x=T_j\sin^2(\theta/2)$, 
we obtain 
\begin{equation}
\label{ithrnseries}
  I(\theta)R_N=\frac{\pi\Delta_0}{2e}\left[
  1 + \sum_{n=1}^{\infty}A_n\frac{\Sigma_{n+1}}{\Sigma_1}\sin^{2n}(\theta/2)
  \right]\sin\theta, 
\end{equation}
with
\begin{equation}
A_n=\frac{1}{4^n}{2n\choose{n}}, \ \ \ \ \ \ 
\Sigma_n=\sum_jT_j^n. 
\end{equation}
The two limiting cases, applicable for a~generic mesoscopic Josephson
junction, are the following: 
If $T_j\ll{}1$ for all $j$-s, Eq.\ (\ref{ithrnseries}) reduces to
\begin{equation}
\label{irntunn}
  I(\theta)R_N\simeq
  \frac{\pi\Delta_0}{2e}\sin\theta, 
\end{equation}
defining the tunneling limit.
For a~perfect ballistic system (or Sharvin contact), we have
$T_j=0$ or $1$, so $\Sigma_1=\Sigma_2=\dots$, etc., and
\begin{equation}
\label{irnball}
  I(\theta)R_N=\frac{\pi\Delta_0}{e}
  \sin(\theta/2)\,{\rm sgn}(\cos\theta/2).  
\end{equation}
In turn, the product $I_cR_N$ is bounded by 
$\pi/2\leqslant{}I_cR_Ne/\Delta_0\leqslant{}\pi$.
As a~second parameter quantifying the distance from the tunneling limit, 
one usually defines the skewness of the current-phase
relation,
\begin{equation}
\label{skewdef}
S= \frac{2\theta_c}{\pi}-1, 
\end{equation}
where $\theta_c$ is the phase difference corresponding to the maximum current
($I_c$), such that $0\leqslant{}S\leqslant{}1$.  

Since $A_1\Sigma_2>A_2\Sigma_3>\dots\geqslant{}0$ in Eq.\ (\ref{ithrnseries}),
a~nonzero value of
$\Sigma_2$ implies that $I_cR_N>\frac{\pi}{2}\Delta_0/e$ and $S>0$.
However, the series in Eq.\ (\ref{ithrnseries}) converges rather slowly.
Earlier calculations for $B=0$ and for heavily-doped disk
($|\mu|\gg{}\hbar{}v_F/r_1$) with $r_2/r_1\rightarrow{}1$ showed that
$S_2/S_1\equiv{}1-F=7/8$ \cite{Ryc24} (with the Fano factor $F$), what implies 
$I_cR_Ne/\Delta_0\approx{}1.94411$ and $S\approx{}0.10826$
if only first two terms in Eq.\ (\ref{ithrnseries}) are taken into account; 
including the cumulants up to $\Sigma_4$ (see Ref.\ \cite{Ryc25} for the
relevant values) one obtains $I_cR_Ne/\Delta_0\approx{}2.17736$ and
$S=0.21779$. These values are considerably lower compared to
$I_cR_Ne/\Delta_0=2.42851$ and $S=0.416008$ \cite{Ryc26b} following from the
exact formula, see Eq.\ (\ref{ijoth}) \cite{seriesfoo}.

\begin{table}[t]
\caption{\label{icrnvsr-table}
The product of critical current and normal state resistance ($I_cR_N$), 
following from the approximation assuming incoherent scattering for 
$2r_c\rightarrow{}r_2-r_1$ and different values of $a=r_1/r_2$. 
The skewness of the current-phase relation ($S$) is also given. 
}
\begin{tabular}{ccc}
\hline\hline
$\ a=r_1/r_2\ $ & $\ \ I_cR_N\  [\,\Delta_0/e\,]\ \ $  &  $\ S\ $ \\ \hline
0  & $\pi/2$ & 0 \\
0.01 &  1.67436  &  0.047112  \\
0.02 &  1.70319  &  0.061296  \\
0.05 &  1.74760  &  0.083877  \\
0.1  &  1.78344  &  $\ \ $0.102778$\ \ $  \\
0.2  &  1.81714  &  0.121126  \\
0.3  &  1.83377  &  0.130401  \\
0.4  &  1.84338  &  0.135832  \\
{\smash{\fboxsep=0pt\llap{\rlap{\fbox{\strut\makebox[1.95in]{}}}~}}
\ignorespaces} 
$\!\!\!$0.5  &  1.84930  &  0.139206  \\
0.6  &  1.85302  &  0.141332  \\
0.7  &  1.85532  &  0.142656  \\
0.8  &  1.85667  &  0.143433  \\
0.9  &  1.85735  &  0.143821  \\
1.0  &  1.85754  &  0.143936  \\
\hline\hline
\end{tabular}
\end{table}

\section{Incoherent scattering at magnetic field}
\label{incoscat}
Before calculating the measurable quantities within the formulas 
presented above, we first present the approximation assuming 
the incoherent transport, obtained by generalizing the derivation of
Ref.\ \cite{Ryc24} for the Corbino-Josephson setup.

In the multimode regime ($kr_1\equiv{}|\mu^\star|\gg{}1$), one can consider
well-defined trajectories (see Fig.\ \ref{setupfig}), and the disk symmetry
causes that the incident angles at the two interfaces ($\gamma_1$, $\gamma_2$)
remain constant after multiple scattering.
In turn, the double-contact formula can be applied \cite{Dat97},
\begin{equation}
\label{tjphi12}
  T_j\simeq{}T_{j,\phi} =
  \frac{T_1T_2}{1+R_1R_2 -2\sqrt{R_1R_2}\cos{\phi}},
\end{equation}
where $T_l=2\cos{\gamma_l}/(1+\cos\gamma_l)$ and $R_l\equiv{}1-T_l$, $l=1,2$.
A~phase $\phi$, gained between the scattering events, is assumed
to be random.
Therefore, an arbitrary analytic function of the transmission
probability $f(T_j)$ can be approximated by 
\begin{equation}
  \left\{f(T_j)\right\}_{\rm incoh}=
  \frac{1}{2\pi}\int_{-\pi}^{\pi}d\phi{}\,f(T_{j,\phi}). 
\end{equation}
Next, the summation over $j$ can be approximated by integration over
$u\equiv{}\sin\gamma_1$, leading to
\begin{equation}
\label{sumftju}
  \sum_j{}f(T_j)\approx{}
  2kr_1\,\frac{1}{2}\int_{u_m}^1{}du\,\{f(T)\}_{\rm incoh},  
\end{equation}
where $u_m$ corresponds to the minimal $\gamma_1$ for which a~trajectory
reaches the outer interface. 

The zero-transmission limit, $2r_c\rightarrow{}r_2-r_1$ with
$r_c=\hbar{}k/(eB)$, is parametrized within $\varepsilon>0$ such that
$2r_c=(1+\varepsilon)(r_2-r_1)$.
For $\varepsilon{}\rightarrow{}0$,
$u_c\simeq{}1-\varepsilon{}\left(1+r_2/r_1\right)$, and the integration over
$u$ in Eq.\ (\ref{sumftju}) is replaced by integration over a~variable
\begin{equation}
  \alpha=\frac{1-u}{\varepsilon(1+r_2/r_1)}
  \ \ \ \ \ \
  (1\geqslant{}\alpha\geqslant{}0), 
\end{equation}
leading to further substitutions in Eq.\ (\ref{tjphi12}) 
\begin{align}
  T_1 &\simeq
  2\sqrt{2\alpha\left(1+{\scriptstyle\frac{1}{a}}\right)}\,\varepsilon^{1/2},
  \label{tt1alpeps}
  \\
  T_2 &\simeq
  2\sqrt{2(1-\alpha)\left(1+{a}\right)}\,\varepsilon^{1/2}.
  \label{tt2alpeps}
\end{align}
Altogether, we have
\begin{equation}
\label{sumjftj}
  \sum_j{}f(T_j)\approx{}2kr_1\,\frac{\varepsilon{}(1+\frac{1}{a})}{2}
  \int_0^1{}\,\frac{d\alpha}{2\pi}\int_{-\pi}^{\pi}d\phi{}f(T_{\alpha,\phi}). 
\end{equation}

For the normal-state resistance, see Eq.\ (\ref{rnlan}), we take
$f(T_j)=g_0T_j$, with the conductance quantum $g_0=4e^2/h$, and the
integrations in Eq.\ (\ref{sumjftj}) can be performed analytically, leading
to
\begin{align}
  R_N^{-1} &\approx{} G_{\rm Sharvin}\,
  \sqrt{2}\left(1+{a}^{-1}\right)\,\varepsilon^{3/2} \nonumber \\
  &\times \left\{
  \frac{2\left(1+\sqrt{a}\right)\left(1-3\sqrt{a}+a\right)}{3(1+a)^{3/2}}\right.
  \nonumber \\
  & \ \ \ \ \ \ \ \left. 
  +\frac{2a}{(1+a)^{2}}\,
  \mbox{artanh}\left[\frac{(1+\sqrt{a})\sqrt{1+a}}{1+\sqrt{a}+a}
  \right]
  \right\}, 
\end{align}
where $G_{\rm Sharvin}\equiv{}2g_0kr_1$.
Similar calculations can be performed for $f(T_j)$ being a~positive-integer
power of $T_j$, leading to $\sum_j{}T_j^m\propto{}\varepsilon{}^{3/2}$, 
but the resulting formulas are too long to be presented.
Referring to the expansion given in Eq.\ (\ref{ithrnseries}) again, 
for $a=1$, we obtain  
$\Sigma_2/\Sigma_1\approx{}0.450297$,
$\Sigma_3/\Sigma_1\approx{}0.310312$, and $\Sigma_4/\Sigma_1\approx{}0.240775$. 
This leads to 
$I_cR_N\approx{}1.75646\,\Delta_0/e$ and $S\approx{}0.06325$ in case
the terms up to $n=1$ are taken into account, or 
$I_cR_N\approx{}1.8271\,\Delta_0/e$ and $S\approx{}0.1110$, if
the terms up to $n=3$ are included. 

The current-phase relation $I(\theta)R_N$, obtained by applying the
approximation as in Eq.\ (\ref{sumjftj}) [with the substitutions given
by Eqs.\ (\ref{tjphi12}), (\ref{tt1alpeps}), and (\ref{tt2alpeps})] 
for Eqs.\ (\ref{ijoth}) and ((\ref{rnlan})) and performing the 
integrations numerically for $a=1$ is depicted in Fig.\ \ref{setupfig}(c).
The corresponding curve [blue line] is apparently closer to the tunneling
limit given by Eq.\ (\ref{irntunn}) [dashed black] than the $B=0$ curve
[red line] taken from Ref.\ \cite{Ryc26b}, yet a~considerable deviation
is still visible in the $2r_c\rightarrow{}r_2-r_1$ limit considered
in the present paper. 

Values of the product $I_cR_N$ --- for different values of the radii ratio $a$
--- are listed in Table~\ref{icrnvsr-table}, where we have also specified the
corresponding values of the skewness $S$. 
In contrast to the similar results for $B=0$, see Ref.\ \cite{Ryc26b},
for which the dependence of $a$ was very weak, we now observe clear
evolution from the tunneling-limit values, approached for $a\rightarrow{}0$, 
towards the nontrivial (neither tunneling nor ballistic) values occurring for
$a\rightarrow{}1$.

Since the characteristics in Table~\ref{icrnvsr-table} saturates starting
from about $a\gtrsim{}0.5$, also taking into account the experimental
limitations \cite{Pol18,Kam21}, we set $a=0.5$ for the numerical
analysis presented in the remaining parts of the paper.

\begin{figure}[t]
\includegraphics[width=\linewidth]{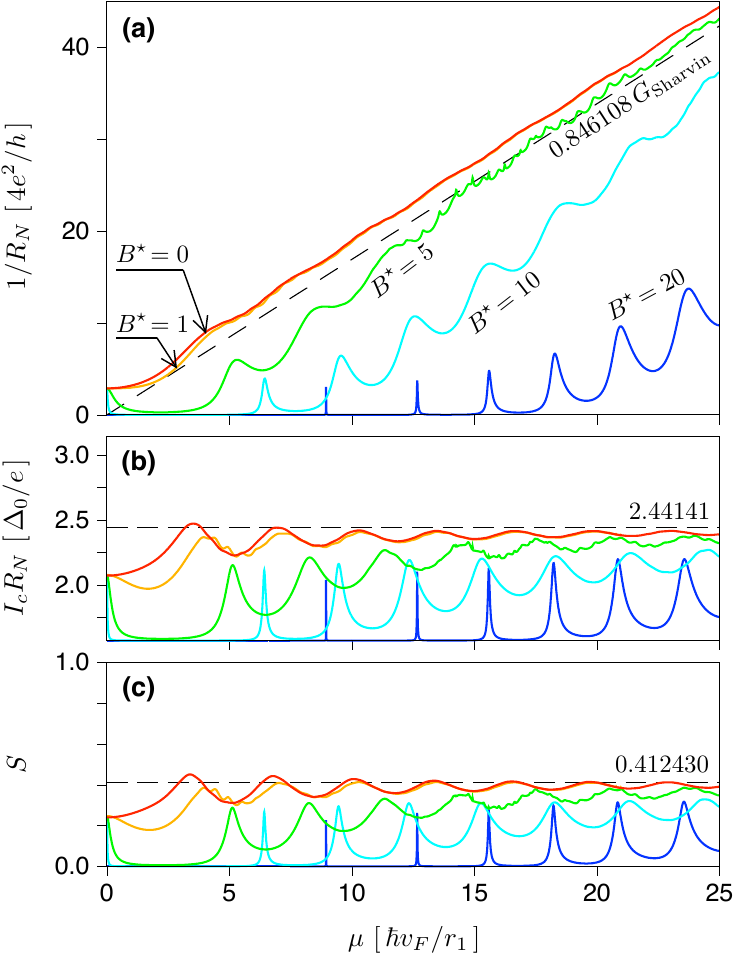}
\caption{ \label{gicsfig}
(a) Normal-state conductance $1/R_N$, (b) product $I_cR_N$, and
(c) skewness of the current-phase relation $S$ as functions of the
chemical potential for different magnetic fields,
$B^\star\equiv{}eBr_1^2/2\hbar$, specified for each line in (a); same
line-color encoding is used in (a)--(c). 
The radii ratio is $r_2/r_1=2$ ($a=0.5$). 
Dashed-black lines correspond to the $B\rightarrow{}0$,
$|\mu|\gg{}\hbar{}v_F/r_1$ limit (see Ref.\ \cite{Ryc26b}). 
}
\end{figure}

\begin{figure}[t]
\includegraphics[width=\linewidth]{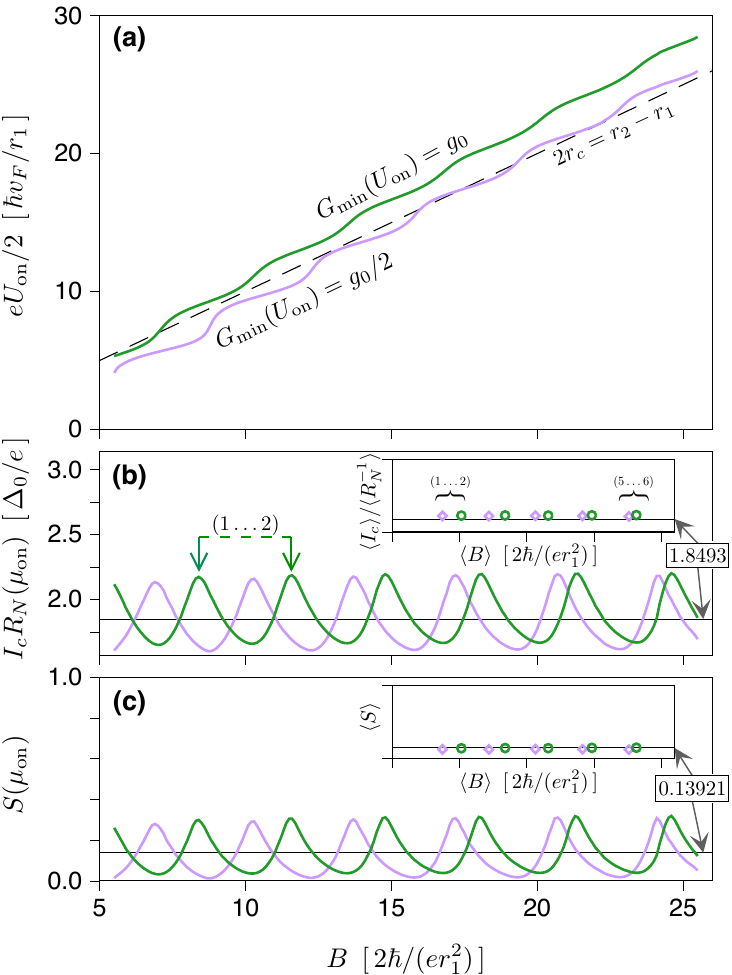}
\caption{ \label{uonicsfig}
  (a) Chemical potential corresponding to the activation voltage
  ($\mu_{\rm on}\equiv{}eU_{\rm on}/2$) for the normal state displayed as
  function of the magnetic field for the two selected threshold conductances:
  $G_{\rm min}(U_{\rm on})=\eta{}g_0$, with $\eta=0.5$, $1$ [solid lines].
  Additional line [dashed] marks $2r_c=r_2-r_1$ (or $\mu_{\rm on}^\star=B^\star$).
  (b) Product $I_cR_N$ and (c) skewness $S$ calculated at $\mu=\mu_{\rm on}$
  [solid lines]. Thin horizontal lines in (b,c) mark the predictions
  for incoherent scattering (see also Table~\ref{icrnvsr-table}). 
  Insets in (b,c) depict the averages taken for the consecutive intervals 
  between the maxima, i.e., ($1\dots{}2$), \dots, ($5\dots{}6$). 
}
\end{figure}

\section{Results and discussion}
\label{resdis}
In this section, we present the results obtained from the exact formula
for transmission probabilities $T_j$, see Eq.\ (\ref{tjbnon}), by performing
the summations in Eqs.\ (\ref{ijoth}) and (\ref{rnlan}) numerically.
The comparison with the approximation assuming incoherent scattering
(see above) will be given then. 

In Fig.\ \ref{gicsfig}, we display the normal-state conductance $1/R_N$,
the product $I_cR_N$, and the skewness $S$ as functions of the
chemical potential $\mu$, for a~fixed $a\equiv{}r_1/r_2=0.5$ and selected
values of the magnetic field [color solid lines in Figs.\
\ref{gicsfig}(a)--(c)].
For the chosen radii ratio the flux quantization, see Eq.\ (\ref{bsquant}),
implies $B^\star\equiv{}eBr_1^2/(2\hbar)=l/6$, with $l$-integer.
The condition under which classical trajectory connects the
electrodes, $2r_c>r_2-r_1$, can be expressed in the dimensionless units of
Eq.\ (\ref{unitdefs}) as follows
\begin{equation}
  \mu^\star > \mu_c^\star\equiv{}\frac{aB^\star}{1-a}=B^\star
  \ \ \ \
  (\text{for }\ a=0.5). 
\end{equation}
Apparently, the conductance is suppressed for $\mu^\star<\mu_c$, see
Fig.\ \ref{gicsfig}(a), except from narrow resonances with Landau levels
(LLs), located near
$\mu_{nLL}^\star=\mbox{sgn}\,n\sqrt{4|n|B^\star}\,$ (with $n$-integer).
Similar resonances appear (if $\mu^\star<\mu_c^\star$) for the remaining
quantities, see Figs.\ \ref{gicsfig}(b) and \ref{gicsfig}(c). 
Also, all the quantities considered approach their multimode
limits calculated in Ref.\ \cite{Ryc26b} [black dashed lines] for
$\mu^\star\gg{}\mu_c^\star$.

In order to reduce the effect of resonances with LLs, and make it possible
to compare the numerical results presented in the above with approximations
given in Sec.\ \ref{incoscat}, we now consider the finite-voltage conductance
in the normal state,
\begin{equation}
\label{guefflan}
  G(\mu,U_{\rm eff}) = \frac{I_N}{U_{\rm eff}}=
  \frac{g_0}{U_{\rm eff}}
  \int_{\mu-eU_{\rm eff}}^{\mu}\!d{}E\,\sum_jT_j(E), 
\end{equation}
where $I_N$ is the normal current, $\mu-eU_{\rm eff}$ and $\mu$ are chemical
potentials in the electrodes, and $T_j(E)$ are given by Eq.\ (\ref{tjbnon}).
(For $U_{\rm eff}\rightarrow{}0$, the linear-response limit given by Eq.\
(\ref{rnlan}) is reproduced.) 
Due to the particle-hole symmetry, $G_{\rm min}(U_{\rm eff})$ --- 
the conditional minimum of $G(\mu,U_{\rm eff})$ for a~fixed $U_{\rm eff}$ ---
corresponds to $\mu_{\rm min}=eU_{\rm eff}/2$, provided that the voltage is of
the order of $U_{\rm eff}\sim{}2\mu_c/e$ (or larger) such that all
well-pronounced LLs are contained within the
$-eU_{\rm eff}/2\dots{}eU_{\rm eff}/2$ energy range. 

Next, we define the activation voltage $U_{\rm on}$, such that
$G_{\rm min}(U_{\rm on})=\eta{}g_0$, choosing the parameter $\eta=0.5$, $1$.
In Fig.\ \ref{uonicsfig}(a), we display the corresponding chemical potential
$\mu_{\rm on}\equiv{}eU_{\rm on}/2$ versus the magnetic field [color solid lines],
together with the line corresponding to the $2r_c=r_2-r_1$ condition
(being equivalent to $\mu_{\rm on}^\star=B^\star$ in the dimensionless units)
[black dashed line]. 

Since the critical current $I_c$ and the skewness $S$ are defined
for the superconducting electrodes, i.e., for $U_{\rm eff}=0$ (the {\em ac}
Josephson effect is beyond the scope of this study), we now take the
value of $\mu_{\rm on}\equiv{}eU_{\rm on}/2$ determined in the previous step
and display --- in Figs.\ \ref{uonicsfig}(b,c) ---
the corresponding product $I_cR_N$ and $S$ as functions of $B$
[color solid lines], together with the predictions for incoherent scattering
also given in Table~\ref{icrnvsr-table}) [black lines]. 
We further notice that the flux quantization implying that $B^\star$ is
the multiplicity of $\frac{1}{6}$, see Eq.\ (\ref{bsquant}), allows
one to visualize the data with continuous lines for the range of
$B^\star\leqslant{}25.5$ used in Fig.\ \ref{uonicsfig}.

All datasets visualized in Fig.\ \ref{uonicsfig} exhibit some periodic
modulation, which we attribute to the resonances with LLs.
In particular, supposing that position of LL coincides with the condition
$2r_c=r_2-r_1$, i.e., $\mu_{nLL}=\mu_c(B)$ for some $n$, we immediately obtain
that $B^\star=B_{n}^\star\equiv{}4n(a^{-1}-1)$, or $B^\star=4n$ for $a=0.5$. 
Although the level broadening, and some energy shift, is significant if
$\mu_{nLL}\approx\mu_c(B)$ (see Fig.\ \ref{gicsfig}), and the above
consideration is only approximate, the period of $\Delta{}B^\star=4$
provides is very close to the actual distances between maxima visible
in Figs.\ \ref{uonicsfig}(b,c). 

A~direct comparison with the predictions for incoherent scattering
(see Table~\ref{icrnvsr-table}) is possible after the influence of the 
above-described resonances with LLs is minimized by taking the average,
independently for each oscillation period, which is defined as the interval
between neighboring maxima for the quantities displayed in
Figs.\ \ref{uonicsfig}(b,c).
The results are presented in the insets, and show approximately $1\%$
agreement with the predictions given in Table~\ref{icrnvsr-table}.

\section{Conclusions}
\label{conclu}
We have pointed out that the current-phase relation for the Corbino-Josephson
junction in graphene subjected to perpendicular magnetic field blocking the
current is significantly different from the sine function characterizing
the tunneling Josephson junction. 
Mathematically, the effect we describe is related to the sub-Poissonian
shot noise (characterized by the Fano factor $F\approx{}0.55$) \cite{Ryc24}
predicted for the same setup with normal metallic leads; however,
the series expressing the  Josephson current via charge-transfer cumulants
converges rather slowly, so the additional computations,
involving both the exact mode-matching analysis and the approximation
assuming incoherent scattering of Dirac fermions between the two circular
sample-lead interfaces. 

Our result complements the list of graphene-specific charge- and
energy-transport phenomena, including the sub-Poissonian shot noise
\cite{Two06}, the mesoscopic Josephson effect at zero (or low) field
\cite{Tit06,Tit07,Hee07,Eng16,Nan17}, anomalous thermoelectric properties
\cite{Sus18,Jay21,Ryc21a,YTu23}, or nonstandard conductance fluctuations
\cite{Hor09,Pal12}.
We hope that recent progress in fabrication of Corbino-Josephson junctions
on the surface of topological insulators \cite{Zha22} and in graphene 
\cite{Par26} would make it possible to confirm experimentally our results
soon.

\vspace{1em}

\section*{Acknowledgments}
The work was partly completed during a sabbatical granted by the Jagiellonian
University in the summer semester of 2024/25. 
We gratefully acknowledge Polish supercomputing infrastructure PLGrid
(HPC Center: ACK Cyfronet AGH) for providing computer facilities and support
within computational grants Nos.\
PLG/2025/018544 (partly) and PLG/2025/018379.




\begin{thebibliography}{}
\bibitem{seriesfoo}
At the Dirac point, $\mu=0$, the convergene is comparably slow, 
namely, we get
$I_cR_Ne/\Delta_0\approx{}1.85084$ and $S\approx{}0.08779$ 
when taking first two terms in the expansion given
by Eq.\ (\ref{ithrnseries}), or
$I_cR_Ne/\Delta_0\approx{}1.98983$ and $S\approx{}0.16667$
when taking the four terms, to be compared with
$I_cR_Ne/\Delta_0\approx{}2.08207$ and $S\approx{}0.25480$
following from the full formula in Eq.\ (\ref{ijoth}).
\end{thebibliography}
\end{document}